\documentclass[10pt]{forma}
\input epsf.tex 
\newcommand{\mincir}{\raise -2.truept\hbox{\rlap{\hbox{$\sim$}}\raise5.truept 
\hbox{$?$}\ }} 
\newcommand{\gr}{\kern 2pt\hbox{}^\circ{\kern -2pt K}} 
\newcommand{\magcir}{\raise -2.truept\hbox{\rlap{\hbox{$\sim$}}\raise5.truept 
\hbox{$?$}\ }} 
 
\newcommand{\Om}{\Omega} 
\newcommand{\be}{\begin{equation}} 
\newcommand{\ee}{\end{equation}} 
\newcommand{\bea}{\begin{eqnarray}} 
\newcommand{\eea}{\end{eqnarray}}

\newcommand{\etal}{{et al.}} 
\newcommand{\apj}{//Astrophys. J.} 
\newcommand{\aj}{//Astron. J.} 
 
\newcommand{\mnras}{//Mon. Not. Roy. Astron. Soc.} 
\newcommand{\aap}{//Astron. and Astrophys.}

\def\be{\begin{equation}}
\def\ee{\end{equation}}

\def\bea{\begin{eqnarray}}
\def\eea{\end{eqnarray}}

\def\etal{{\it et al.}}
\begin{document}

\hspace{1.0cm} \parbox{15.0cm}{

\baselineskip = 15pt

\noindent {\bf 
Cosmological~ parameters~ from~ observational~ data \\
on~ the~ large~ scale~ structure~ of~ the~ Universe
}

\bigskip
\bigskip

\noindent {\bf B.~Novosyadlyj$^{(1)}$, R.~Durrer$^{(2)}$ and S.~Apunevych$^{(1)}$}

\bigskip

\baselineskip = 9.5pt

\noindent {\small \copyright~2000}

\smallskip

$^1$\noindent {\small {\it  Astronomical Observatory of National University of L'viv\\
Kyryla i Mephodia St. 8, 79005 L'viv, Ukraine}} \\
\noindent {\small {\it e-mail:}} {\tt novos@astro.franko.lviv.ua, apus@astro.franko.lviv.ua}

\smallskip

$^2$\noindent {\small {\it Department de Physique Th\'eorique, Universit\'e de Gen\`eve \\
Quai Ernest Ansermet 24, CH-1211 Gen\`eve 4, Switzerland}}\\
\noindent {\small {\it e-mail:}} {\tt ruth.durrer@physics.unige.ch}
\baselineskip = 9.5pt \medskip

\medskip \hrule \medskip

\noindent The observational data on the large scale structure (LSS) of the Universe 
are used to determine cosmological parameters within the class of adiabatic 
inflationary models. We show that a mixed dark matter model with 
 cosmological constant ($\Lambda$MDM  model) and  parameters   
$\Omega_m=0.37^{+0.25}_{-0.15}$,     
$\Omega_{\Lambda}=0.69^{+0.15}_{-0.20}$,     
$\Omega_{\nu}=0.03^{+0.07}_{-0.03}$,    $N_{\nu}=1$, 
$\Omega_b=0.037^{+0.033}_{-0.018}$,  
$n_s=1.02^{+0.09}_{-0.10}$,     
$h=0.71^{+0.22}_{-0.19}$,     
$b_{cl}=2.4^{+0.7}_{-0.7}$ 
 (1$\sigma$ confidence limits)     
matches  observational data on  LSS, the nucleosynthesis constraint, direct 
measurements of the Hubble constant, the  high redshift supernova type Ia  
results and the recent measurements of the location and amplitude of 
the first acoustic peak in the CMB anisotropy power spectrum. 
The best model is $\Lambda$ dominated (65\% of the total energy density) 
and has slightly positive curvature, $\Omega=1.06$. The  
clustered  matter consists in 8\%  massive neutrinos, 10\%  baryons 
and 82\%  cold dark matter (CDM).  It is shown  that the LSS observations  
together with  the Boomerang (+MAXIMA-1) data on the first acoustic 
peak  rule out zero-$\Lambda$ models at more than $2\sigma$ 
confidence level.

\medskip \hrule \medskip

}
\vspace{1.0cm}

\baselineskip = 11.2pt

\noindent {\small {\bf INTRODUCTION}}
\medskip

\noindent The  comparison of recent experimental 
data on the large scale structure of the Universe 
with theoretical predictions of inflationary cosmology have shown 
since quite some time that the simplest cold dark matter (CDM) model 
is ruled out and we have to allow for a wider set of parameters to fit 
all observational data on the state and history of our Universe.  
These include spatial curvature ($\Omega_k$), a  
cosmological constant ($\Omega_{\Lambda}$), the Hubble parameter 
($h\equiv H_0/(100$km/s/Mpc$)$), the energy density of baryonic matter 
($\Omega_b$),  cold dark matter ($\Omega_{cdm}$), the number of 
species of massive neutrinos ($N_{\nu}$) and their density ($\Omega_{\nu}$), 
the amplitude of the power spectra of primordial perturbations in 
scalar ($A_s$) and tensor ($A_t$) modes and the corresponding 
power-law indices  ($n_s$ and $n_t$), and the  optical depth to early 
reionization ($\tau$). Constraining  this multidimensional parameter 
space, determining the true values of fundamental cosmological 
parameters, the nature and content of the matter which fills our 
Universe is an important and exciting  problem of cosmology which has 
now become feasible due to the enormous progress in cosmological observations
(short reviews and references see in \cite{nov00a,dur00}).

The goal of this paper is to determine  cosmological parameters of the 
sub-class of models without tensor mode and no early reionization on 
the basis of LSS data related to different scales and different 
redshifts.  We test $\Lambda$MDM models with non-zero 
curvature. Since  the sum $\Omega_k +\Omega_{\Lambda}+\Omega_m =1$ 
according to Friedmann's equation, we treat $\Omega_{\Lambda}$ and 
$\Omega_m$  as free parameters. Furthermore,  we use the data on the 
location and amplitude of the first acoustic peak determined from 
the most accurate recent measurements of the CMB power spectrum. 
We also use the  SNIa constraint for comparison. 
 
The outline of the paper is as follows: in Sect.~1  
we describe the experimental data set which is used here
 and the method employed to determine cosmological parameters. 
In Sect.~2 we present and discuss our results. 
Our conclusions are presented in Sect.~3.  
 
\section{The experimental data set and method of calculation} 
 
Our approach is based on the comparison of the observational data on the 
structure of the Universe over a wide range of scales with theoretical 
predictions from the power spectrum of small (linear) density fluctuations.
The form of the spectrum strongly depends on the 
cosmological parameters $\Omega_m$, $\Omega_b$, $\Omega_{\nu}$, $N_{\nu}$,
$h$ and $n_s$. If its amplitude at one given scale is fixed by some
observational data then predictions for observations on other scales
can be calculated and compared with 
corresponding observational data. Minimization of the quadratic differences 
between the theoretical and observational values divided by the observational 
errors, $\chi^2$, determines the best-fit values of the 
above mentioned cosmological parameters. For this we use the following
observational data set:

1. The location $\tilde\ell_p=197\pm6$ and amplitude $\tilde A_p=69\pm8\mu \rm K$
of the first acoustic  peak in the angular power spectrum of the CMB 
temperature fluctuations deduced from CMB map obtained in Boomerang 
experiment \cite{ber00} (these data are sensitive to the amplitude and form of the
initial power spectrum in the range $\approx 200h^{-1}$Mpc);

2. The power spectrum of density fluctuations of Abell-ACO clusters ($\tilde P_{A+ACO}(k_j)$)
obtained from their space distribution by  \cite{ret97}
(scale range $10-100h^{-1}$Mpc);

3. The constraint for the amplitude of the fluctuation power spectrum on 
$\approx 10h^{-1}$Mpc scale derived from a recent optical determination of the 
mass function of nearby galaxy clusters \cite{gir98} gives 
$\tilde\sigma_{8}\Omega_m^{\alpha_1}=0.60\pm 0.04$ 
where $\alpha_1=0.46-0.09\Omega_m$ for flat low-density models and 
$\alpha_1=0.48-0.17\Omega_m$ for open models (at the 90\% C.L.);
 
4. The constraint for the amplitude of the fluctuation power spectrum on 
$\approx 10h^{-1}$Mpc scale derived from the observed evolution of the galaxy 
cluster X-ray temperature distribution function between $z=0.05$ and $z=0.32$
 derived by \cite{via99} 
$\tilde\sigma_8\Omega_m^{\alpha_2}=0.56\pm0.19\Omega_m^{0.1\lg\Omega_m+\alpha_2}, 
\;\;\;\alpha_2=0.34$ 
for open models and 
$\tilde\sigma_8\Omega_m^{\alpha_2}=0.56\pm0.19\Omega_m^{0.2\lg\Omega_m+\alpha_2}, 
\;\;\;\alpha_2=0.47$ 
for flat models (both with 95\% confidence limits);

5. The constraint for the amplitude of the fluctuation power spectrum on 
$\approx 10h^{-1}$Mpc scale derived from the existence of three very massive 
clusters of galaxies observed 
so far at $z>0.5$ \cite{bah98} 
$\tilde \sigma_8\Omega_m^{\alpha_3}=0.8\pm 0.1\;$, 
where $\alpha_3=0.24$ for open models and $\alpha_3=0.29$ for flat 
models;
  
6. The constraint on the amplitude of the linear power spectrum of 
density fluctuations in our vicinity obtained from the study of  
bulk flows of galaxies in sphere of radius 
$50h^{-1}$Mpc  $\tilde V_{50}=(375\pm 85) {\rm km/s}$ 
\cite{kol97};

7. The constraint for the amplitude of the fluctuation power spectrum on 
$0.1-1h^{-1}$Mpc scale and $z=3$ derived from the Ly-$\alpha$ absorption lines seen in 
quasar spectra $ 1.6<\tilde \sigma_{F}(z=3)<2.6$ (95\% C.L.) at 
$k_{F}\approx 38\Omega_m^{1/2}h/{\rm Mpc}$ \cite{gn98,ric99,gn99};
 
8. The constraints for the amplitude and inclination of the initial power 
spectrum on $\approx 1h^{-1}$Mpc scale and $z=2.5$ scale derived by Croft \etal (1998)
from the Ly-$\alpha$ forest of quasar absorption lines 
$\tilde \Delta_{\rho}^2(k_p)\equiv k_p^3P(k_p)/2\pi^2=0.57\pm 0.26,$ 
$\tilde n_p\equiv {\Delta \log\;P(k)\over \Delta \log\;k}\mid 
_{k_p}=-2.25\pm 0.18,$ $k_{p}=1.5\Om_m^{1/2}h/$Mpc,   at (95\% CL);
 
9. The data on the direct measurements of the Hubble constant 
$\tilde h=0.65\pm 0.10$ which is a compromise between results obtained 
by two groups \cite{tam97} and \cite{mad98}; 

10. The nucleosynthesis constraint on the baryon density derived from a 
observational content of inter galactic deuterium 
$\widetilde{\Omega_bh^2} = 0.019\pm 0.0024$ (95\% C.L.) 
given by \cite{bur99};

11. The constraint on the matter and vacuum (cosmological constant) 
energy density derived from the distance measurements of super novae 
of type Ia (SNIa)~\cite{rie98,per98,per99} in the form
$\widetilde{[\Omega_m-0.75\Omega_{\Lambda}]}=-0.25\pm0.125~.$

One of the main ingredients for the solution for our search problem 
is a reasonably fast and accurate determination of the  linear 
transfer function for dark matter clustering 
which depends on the cosmological parameters. We use accurate analytical 
approximations of the MDM transfer function $T(k;z)$ depending on 
the parameters $\Omega_m$, $\Omega_b$, $\Omega_{\nu}$, $N_{\nu}$ and 
$h$ given by \cite{eh3}. The 
linear power spectrum of matter density fluctuations
$P(k;z)=A_sk^{n_s}T^2(k;z)D_1^2(z)/D_1^2(0)$, 
where $A_s$ is the normalization constant for scalar perturbations and 
$D_1(z)$ is the linear growth factor. 
We normalize the spectra to the 4-year COBE data \cite{ben96}
which determine the amplitude of density perturbation at horizon 
scale, $\delta_h$ \cite{lid96,bun97}. Therefore, each model will match
the COBE data by construction. The normalization constant $A_s$ is 
then given by $A_s=2\pi^{2}\delta_{h}^{2}(3000{\rm Mpc}/h)^{3+n_s}$. 
 
The Abell-ACO power spectrum is related to the matter power 
spectrum at $z=0$, $P(k;0)$, by the cluster biasing parameter $b_{cl}$:
$P_{A+ACO}(k)=b_{cl}^{2}P(k;0)$.
We assume scale-independent linear bias as free parameter of which best-fit
values will be determine join with other cosmological parameters.
 
The dependence of the position and amplitude of the first acoustic 
peak in the CMB power spectrum on cosmological 
parameters $n_s$, $h$, $\Omega_b$, $\Omega_{cdm}$ and 
$\Omega_{\Lambda}$ can be calculated using the analytical approximation 
given in \cite{efs99} which has been  extended to  models with non-zero 
curvature  ($\Omega_k\equiv1-\Omega_m-\Omega_{\Lambda}\ne0$) in 
\cite{dur00}. Its accuracy in the parameter ranges 
which we consider is better then 5\%.

The theoretical values of the other experimental constraints are
calculated as described in \cite{nov00a}. There
one can also find  tests of the method.

\section{Results and Discussion}

The determination of the parameters  
$\Omega_m$, $\Omega_{\Lambda}$, $\Omega_{\nu}$, $N_{\nu}$, $\Omega_b$,  
$h$, $n_s$ and $b_{cl}$ by the Levenberg-Marquardt $\chi^2$ 
minimization method can be realized in the following way: we vary the 
set of parameters  
$\Omega_m$, $\Omega_{\Lambda}$, $\Omega_{\nu}$, $\Omega_b$,  
$h$, $n_s$ and $b_{cl}$ with fixed $N_{\nu}$ and find the 
minimum of 
$\chi^2$. Since $N_{\nu}$ is discrete we repeat this 
procedure three times for $N_{\nu}$=1, 2, and 3.  The lowest of the 
three minima is the minimum of $\chi^2$ for the 
complete set of free parameters.  
Hence, we have seven free parameters. 
The formal number of observational points is 25 but, as it was shown
in \cite{nov00a}, the 13 points of the cluster power spectrum  can be
described by just 3 degrees of freedom, so that the  
maximal number of truly independent measurements is 15. Therefore,  
the number of degrees of 
freedom for our search procedure is $N_F= N_{\rm exp}-N_{\rm par}= 8$  
if all observational points are used. In order to investigate to what extent the 
LSS constraints on fundamental parameters match the constraints implied by 
SNIa \cite{per99} we have determined all 8 parameters without (i) and  
with (ii) the SNIa constraint. In the case without LSS constraints (iii) 
the number of residual experimental points equals with number of free parameters 
\footnote{$\Omega_{\nu}$ and $b_{cl}$ are not determined in this case} ($N_F=0$). 
The results are presented in the Table~\ref{rez1}.  
\begin{table}[h] 
\caption{Cosmological parameters determined from the different set 
of observational data: (i) - LSS, $\tilde\ell_p$, $\tilde A_p$, $\tilde h$, $\widetilde{\Omega_bh^2}$,
(ii) - LSS, $\tilde\ell_p$, $\tilde A_p$, $\tilde h$, $\widetilde{\Omega_bh^2}$, SNIa constraint,
(iii) - $\tilde\ell_p$, $\tilde A_p$, $\tilde h$, $\widetilde{\Omega_bh^2}$, SNIa constraint.}  
\begin{center} 
\def\onerule{\noalign{\medskip\hrule\medskip}} 
\begin{tabular}{|ccccccccc|} 
\hline 
&&&&&&&&\\ 
Data set&$\chi^2_{min}/N_F$& $\Omega_m$& $\Omega_{\Lambda}$&$\Omega_{\nu}$& $\Omega_b$  &$n_s$   & $h$ &$b_{cl}$ \\ [4pt] 
\hline 
&&&&&&&&\\
(i)&5.90/7&$0.37^{+0.25}_{-0.15}$&$0.69^{+0.15}_{-0.20}$&$0.03^{+0.07}_{-0.03}$&$0.037^{+0.033}_{-0.018}$&$1.02^{+0.09}_{-0.10}$&$0.71^{+0.22}_{-0.19}$&$2.4^{+0.7}_{-0.6}$\\  
&&&&&&&&\\
(ii)&6.02/8&$0.32^{+0.20}_{-0.11}$&$0.75^{+0.10}_{-0.19}$&$0.0^{+0.09}_{-0.0}$&$0.038^{+0.033}_{-0.019}$&$1.00^{+0.13}_{-0.10}$&$0.70^{+0.28}_{-0.18}$&$2.2^{+0.8}_{-0.5}$\\
&&&&&&&&\\
(iii)&0/0&0.33$\pm$0.07&0.77$\pm$0.08&-&0.045$\pm$0.014&0.96$\pm$0.06&0.65$\pm$0.1&-\\
\hline  
\end{tabular}  
\end{center} 
\label{rez1}
\end{table} 
Note, that for all models $\chi^2_{min}$ is in the range 
$N_F-\sqrt{2N_F}\le \chi^2_{\min}\le N_F+\sqrt{2N_F}$ which is 
expected for a Gaussian distribution of $N_F$ 
degrees of freedom. This means that the cosmological paradigm which 
has been  assumed is in agreement with the data. 
Including the MAXIMA-1 \cite{han00} data 
into the determination of the first acoustic peak does not change the results 
essentially (see \cite{dur00}).  
The errors in the best-fit parameters presented in Table~\ref{rez1} 
for cases (i) and (ii) are obtained by maximizing the (Gaussian) 68\%
confidence contours over all other parameters, the errors for the case (iii)
are the square roots of the  diagonal elements of the covariance 
matrix.  
 
The 
model with one sort of massive neutrinos provides the best fit to 
the data, $\chi^2_{min}=5.9$.  
However,  there is 
only a marginal difference in $\chi^2_{min}$ for $N_\nu =1,2,3$.   
With the given accuracy of the data we cannot conclude 
whether  massive neutrinos are present 
at all, and if yes what number of degrees of freedom is favored.  
We summarize, that the 
considered observational data on LSS of the Universe can be 
explained by a $\Lambda$MDM inflationary model with a scale invariant 
spectrum of scalar perturbations and small positive curvature. 
 
Including of the SNIa constraint into the experimental data set 
 decreases $\Omega_m$, increases $\Omega_{\Lambda}$ slightly and 
prefers $\Omega_{\nu}\approx0$, a $\Lambda$CDM model. Excluding the
LSS constraints (items 2-8 in sect.2) does not significantly change the best-fit
values (of those parameters which are determined), which
demonstrates nicely the concordance of different experimental data sets
and their theoretical interpretation.
 
In the cases (i) and (ii), the calculated characteristics of the LSS 
are within the $1\sigma$ error bars of the observed values. 
The predicted age of the Universe agrees well with recent  
determinations of the age of globular clusters.

 Comparing the results obtained without and with the SNIa constraint, 
we conclude that the values of the fundamental cosmological parameters 
$\Omega_m$, $\Omega_{\Lambda}$ and  
$\Omega_k$  determined by the observations of large 
scale structure match the SNIa test very well. This can be interpreted 
as independent support of the SNIa result in the framework of the 
standard cosmological paradigm. However, 
in order to elucidate how LSS data constraint cosmological parameters, 
we analyze further only the model obtained without the SNIa constraint. 
  
The best fit values of cosmological parameters determined by LSS 
characteristics are presented in the 1-st row of Table~\ref{rez1}.  
The best-fit CDM density parameter is $\Omega_{cdm} = 0.30$ and 
$\Omega_{k}=-0.06$, slightly positive curvature.

The value of the Hubble constant is close to the result by 
\cite{mad98} and \cite{mou00}. The spectral index coincides with 
the prediction of the simplest inflationary scenario, it is close 
to unity. 
The neutrino matter density $\Omega_{\nu}=0.03$ 
corresponds to a neutrino mass of $m_{\nu}=94\Omega_{\nu}h^2\approx1.4$ eV 
but is compatible with 0 within $1\sigma$. 
So, $\Lambda$CDM model ($\Omega_{\nu}=0$) is within the  1$\sigma$
contour of the best-fit $\Lambda$MDM model.  
The 
estimated cluster bias parameter $b_{cl}=2.4$ fixes the amplitude 
of the Abell-ACO power spectrum.

The errors shown in Table~\ref{rez1} (i) define the range of each parameter  
within which by adjusting the remaining parameters a value of  
$\chi^2_{min}\le 11.8$ can be achieved. Of course, the values of the  
remaining parameters always lay within their corresponding 
68\% likelihoods given in Table~\ref{rez1}.  
Models with vanishing $\Lambda$ are outside of this marginalized 
$1 \sigma$ range of the best-fit model determined by the LSS 
observational characteristics used here even without the SNIa 
constraint. Indeed, if we set $\Omega_{\Lambda}=0$ as fixed parameter 
and determine  the remaining parameters in the usual way, the minimal 
value of  $\chi^2$ is $\chi^2\approx24$ with the following values for 
the other parameters: $\Omega_m=1.15$, $\Omega_{\nu}=0.22$, 
$N_{\nu}=3$, $\Omega_b=0.087$, $n_s=0.95$, $h=0.47$, $b_{cl}=3.7$ 
($\sigma_8=0.60$)). This model is outside the  2$\sigma$ confidence 
contour of  the best-fit model (i). The experimental data  which disagree most with 
$\Lambda=0$ are the data on the first acoustic peak. If we exclude it 
from the experimental data set, $\chi^2_{min}\approx5.8$ for an open model 
with the following best-fit parameters: $\Omega_m=0.48$, $\Omega_{\nu}=0.12$, 
$N_{\nu}=1$,  $\Omega_b=0.047$, $n_s=1.3$, $h=0.64$, $b_{cl}=2.5$  
($\sigma_8=0.82$)).  
This model is within the  1$\sigma$ confidence contour of the best-fit 
$\Lambda$MDM model obtained without data on the first acoustic peak. 
The reason for this 
behavior is clear: the position of the 'kink' in the matter power 
spectrum at large scales demands a 'shape parameter' 
$\Gamma=\Omega_mh^2 \sim 0.25$ which can be achieved either by 
choosing an open model or allowing for a cosmological constant. The 
position of the acoustic peak which demands an approximately flat model then 
closes the first possibility. 
 
Results change only 
slightly if instead of the Boomerang data we use Boomerang+MAXIMA-1. 
Hence, we  can conclude that the LSS observational  
characteristics together with the Boomerang (+MAXIMA-1) data on the 
first acoustic peak already rule out  
zero-$\Lambda$ models at more than 95\% C.L.  and actually demand a 
cosmological constant in the same bulk part as direct measurements. We 
consider this a non-trivial consistency check! 
 
Flat $\Lambda$ models in contrary, are inside the 1$\sigma$ 
contour of our best-fit model. Actually, the best fit flat model has  
$\chi^2_{min}\approx8.3$  and the best fit 
parameters $\Omega_m=0.35\pm0.05$, $\Omega_{\Lambda}=0.65\mp0.05$,  
$\Omega_{\nu}=0.04\pm0.02$, $N_{\nu}=1$, $\Omega_b=0.029\pm0.005$,  
$n_s=1.04\pm0.06$, $h=0.81\pm0.06$, $b_{cl}=2.2\pm0.2$  
($\sigma_8=0.96$) are close to our previous~\cite{nov00a} results 
with a somewhat different observational data set. 
It is obvious, that flat zero-$\Lambda$ CDM and MDM models are ruled 
out by the present experimental data set at even higher confidence 
limit than by data without the Boomerang and MAXIMA-1 
measurements in~\cite{nov00a}. 
\footnote{
The more detailed comparison of
results presented here with other parameter estimations one can find
in the accompanying journal paper \cite{dur00}.}

\section{Conclusions} 
 
The main observational characteristics on  LSS together with recent 
data on the amplitude and location  
of the first acoustic peak in the CMB power spectrum, and the 
amplitude of the primordial power spectrum inferred by the COBE DMR 
four year data prefer a $\Lambda$MDM model with the following parameters:  
$\Omega_m=0.37^{+0.25}_{-0.15}$,     
$\Omega_{\Lambda}=0.69^{+0.15}_{-0.20}$,     
$\Omega_{\nu}=0.03^{+0.07}_{-0.03}$,    $N_{\nu}=1$, 
$\Omega_b=0.037^{+0.033}_{-0.018}$,  
$n_s=1.02^{+0.09}_{-0.10}$,     
$h=0.71^{+0.22}_{-0.19}$,     
$b_{cl}=2.4^{+0.7}_{-0.7}$ 
(1$\sigma$ marginalized ranges).        
  
The central values correspond to a slightly closed ($\Omega_k=-0.06$) 
$\Lambda$MDM model with one sort of 1.4eV neutrinos. These neutrinos 
make up about 8\% of the clustered matter, baryons are 10\% 
and the rest (82\%) is in a cold dark matter component. The energy 
density of clustered matter corresponds to only 35\% of the total 
energy density of matter plus vacuum which amounts to $\Omega= 1.06$.

The values for  the matter density $\Omega_m$ and the 
cosmological constant $\Omega_{\Lambda}$ for the best-fit model are 
close to those  deduced from the SNIa test. Including this test in the 
observational data set, results to a somewhat larger value of 
$\Omega_{\Lambda}$  (7\%) and  slightly lowers $\Omega_m$.  
 
The observational characteristics of large scale structure together 
with the  Boomerang (+MAXIMA-1) data on the first acoustic peak rule out  
zero-$\Lambda$ models at more than  $2\sigma$ confidence limit.  
 
\bigskip
\noindent {\small {\bf REFERENCES}}
   

\end{document}